\documentstyle[aps,eclepsf]{revtex}

\begin{document}
\draft

\title{ Theory of the
d.c. Josephson effect in Pb/Sr$_{2}$RuO$_{4}$/Pb}

\author{Masashi Yamashiro and Yukio Tanaka}

\address{Graduate School of Science and Technology,
Niigata University, Ikarashi,
Niigata 950-2181, Japan }

\author{Satoshi Kashiwaya \cite{ETL}}
\address{E. L. Ginzton Laboratory, Stanford University,
Stanford, CA 94305-4085, USA}
\date{\today}
\maketitle
\begin{abstract}
To clarify the origin
of  anomalous behaviors in Pb/Sr$_{2}$RuO$_{4}$/Pb junctions
in terms of the pairing symmetry,
a theory of the d.c. Josephson current in 
$s$-wave superconductor / $p$-wave superconductor /
$s$-wave superconductor junctions is developed.
%We assume two kinds of unitary triplet superconducting
%states with $p$-wave symmetry as the candidates for Sr$_{2}$RuO$_{4}$.
Calculated results on the temperature dependence of the
critical Josephson current exhibit
non-monotonous behaviors when the
thickness of the $p$-wave superconductor
is comparable to the coherence length.
The consistency between present results with recent experimental measurement
supports the possibility
of a unitary $p$-wave pairing state in Sr$_{2}$RuO$_{4}$.
\end{abstract}
\vspace{24pt}

\pacs{}
\widetext
Recently, physical properties on Sr$_{2}$RuO$_{4}$
have been studied vigorously
both in the superconducting states and in the normal state
\cite{maeno2}.
There have been presented several experimental results
as well as theories that support
an unconventional superconducting states
$i$.$e$. triplet $p$-wave pair potential, in
superconducting Sr$_{2}$RuO$_{4}$
\cite{nishizaki,yoshida1,ishida,rice,bakaran,mazin,sigrist,machida}.
In order to identify definitively the pairing symmetry,
phase sensitive measurements utilizing the
Josephson junctions give the most useful information \cite{Ges,van}.
Standing on this view point,
it is an interesting problem to develop a theory of the
Josephson effect for junctions which include triplet
$p$-wave superconductors.
\par
One of the most significant difference of
unconventional superconductors from conventional ones is the
existence of the internal phase in the pair potential.
The important role of the internal phase on the
current-phase relation of the Josephson current
has been pointed out theoretically \cite{sigrist1}.
In a two-dimensional junction configuration with $d$-wave
superconductors,
for a fixed external phase difference
between two superconductors,
the component of the Josephson current becomes either
positive or negative depending on the injection angle of the
quasiparticle \cite{tanaka}.
In addition to this, recent studies have found the
formation of the zero-energy states (ZES) near the surfaces and boundaries
under the influence of the internal phase \cite{hu,tanaka1,kashiwa}.
The critical Josephson current $J_{C}(T)$ via the ZES
\cite{tanaka,barash,samanta97}
is shown to be proportional to $T^{-1}$.
The interplay of above two effects induces a
non-monotonous temperature dependence  of the Josephson current
depending on the orientation of the junctions.
As for  the current-phase relation,
it can be decomposed into
the infinite series of $\sin(n\varphi)$ with integer $n$,
\begin{equation}
J(\varphi)=\sum_{n} \bar{J}_{n} \sin(n\varphi),
\end{equation}
where $\varphi$ is the difference of the external phase
between two superconductors \cite{ishii}.
%In some orientations of the junctions including unconventional
%superconductors,
%Josephson current $J(\varphi)$ is shown to be proportional to $\sin(2\varphi)$
%due to the vanishment of $\bar{J}_{1}$ \cite{tanaka,yip}.
It was revealed in previous papers\cite{yip,pals,fenton} that
$\bar{J}_{1}$ vanishes
in singlet superconductor / triplet superconductor junctions
independent of the orientation of the junctions
if the spin orbit coupling is absent.
This peculiar property originates from the
difference in the parities between the two superconductors.
\par
Recent experiment by Jin reported
a non-monotonous temperature dependence of the
critical Josephson current in Pb/Sr$_{2}$RuO$_{4}$/Pb junctions \cite{jin}.
Since this behavior cannot be understood in terms of
conventional theory of the Josephson effect,
it is interesting to clarify the origin in terms of the influence of the
$p$-wave pair potentials on the Josephson current lying between two
$s$-wave superconductors.
In this paper,
we present a theory of the Josephson current in a 
$s$-wave/ $p$-wave/$s$-wave ($s/p/s$) trilayer superconductor junction
where the transition temperature of
$p$-wave superconductor $T_{p}$ is lower than
that of $s$-wave superconductor $T_{s}$.
%
%We obtain the non-monotonous temperature dependence of the
%Josephson current depending on the  width of the $p$-wave
%superconductor. This result is qualitatively
%consistent with
%recent experiments of Josephson current
%in Pb / Sr$_{2}$RuO$_{4}$ / Pb junction by Jin\cite{jin}.
\par
In the following, we assume an $s/p/s$ junction in the clean limit with
semi-infinite $s$-wave superconductor region ($x<0$ and $x>L$) where the
thickness of the  $p$-wave superconductor is $L$.
The flat interfaces are perpendicular to the $x$-axis,
and are located at $x=0$ and $x=L$.
The barrier potential at the interface is ignored for simplicity.
We also assume that the
Fermi wave number $k_{F}$ and the effective mass $m$ are equal
in the all three regions.
To express the nearly two-dimensional Fermi surface, the $z$ component of
the Fermi momentum $k_{Fz}$ is restricted to the region given by
$-\delta<\sin^{-1}(k_{Fz}/k_{F})<\delta$.
The two-component wave function $\Psi(x)$ of the quasiparticles
is given as the solution of
the Bogoliubov-de Gennes (BdG) equation following the quasi-classical
approximations \cite{bruder}. In this approximation, the
effective pair potentials felt by the quasiparticles  depend
only on the
direction of the motion of the quasiparticles.
The pair potential of the system is assumed to be the
stepwise form\cite{hurd} (see Fig.1)
\[
\Delta_{\sigma\sigma^{\prime}}(x,\theta,\phi)
=
\delta_{\sigma,-\sigma^{\prime}}
(\delta_{\sigma,\uparrow} -\delta_{\sigma,\downarrow})
\Delta_{s}(T)e^{-i\varphi/2}\hspace{12pt} x<0,
\]
\[
\hspace{48pt}=\delta_{\sigma,-\sigma^{\prime}}\Delta_{p}(T,\theta,\phi)
\hspace{16pt} 0<x<L,
\]
\begin{equation}
\hspace{40pt}=\delta_{\sigma,-\sigma^{\prime}}
(\delta_{\sigma,\uparrow} -\delta_{\sigma,\downarrow})
\Delta_{s}(T)e^{i\varphi/2}\hspace{14pt} L<x.
\label{e1}
\end{equation}
In the above,
$\theta$ is the polar angle and $\phi$ is the
azimuthal angle in the $x$-$y$ plane.
For the $p$-wave pair potential, we assume an unitary state
where only $\Delta_{\uparrow\downarrow}(x,\theta,\phi)$
$[=\Delta_{\downarrow\uparrow}(x,\theta,\phi)]$ is non-zero
for simplicity \cite{yip}.
The effective pair potential felt by the
quasiparticles changes its sign depending
on the spin index in the $s$-wave region.
While in the $p$-wave region,
the pair potential has the
same sign regardless of the spins of quasiparticles.
The external
phase difference between the 
$p$-wave superconductor and the $s$-wave
superconductor at each interfaces is $\varphi/2$
as to satisfy the current conservation law \cite{hurd}.
\par
Suppose an electron-like quasiparticle (ELQ) is injected
from the left superconductor  with the injection
angles $\theta$ and $\phi$.
In the present case, the possible reflection process is only
Andreev reflection because of the
absence of the barrier potential at the interfaces.
The coefficients of the Andreev reflection is determined by solving
the BdG equation under the following boundary conditions:
\begin{eqnarray}
\left. \Psi(x)\right|_{x=0_{-}}=
\left. \Psi(x)\right|_{x=0_{+}},\hspace{20pt}
\left. \Psi(x)\right|_{x=L_{-}}=
\left. \Psi(x)\right|_{x=L_{+}}, \\
\left. \frac{d\Psi(x)}{dx}\right|_{x=0_{-}}=
\left. \frac{d\Psi(x)}{dx}\right|_{x=0_{+}},\hspace{20pt}
\left. \frac{d\Psi(x)}{dx}\right|_{x=L_{-}}=
\left. \frac{d\Psi(x)}{dx}\right|_{x=L_{+}}.
\label{e2-1}
\end{eqnarray}
The obtained coefficients depend on the
direction of the spin of the injected ELQ.
For the calculation of Josephson current, we will
extend the formula by
Furusaki and Tsukada \cite{furusaki}  for $s$-wave superconductors.
In this formula, the Josephson current is
expressed by the generalized Andreev coefficients
$\bar{a}_{1(2)}(\varphi,\theta,\phi)$
which are obtained by the analytic
continuation from $E$ to $i\omega_{n}$ in $a_{1(2)}(\varphi,\theta,\phi)$
where $E$ is the energy of the
quasiparticles measured from the Fermi energy
$E_{F}$ and $\omega_{n}=2\pi k_{B}T(n+1/2)$ is the Matsubara frequency.
Then, the Josephson current $J(\varphi)$ is given as
\begin{equation}
eR_{N}J(\varphi)=\Delta_{s}(T)k_{B}T
\displaystyle
\frac{\int_{\pi/2-\delta}^{\pi/2+\delta}\!\int_{-\pi/2}^{\pi/2}
K(\varphi,\theta,\phi) \sin^{2}\theta\cos\phi d\theta d\phi}
{\frac1\pi\int_{\pi/2-\delta}^{\pi/2+\delta}\!\int_{-\pi/2}^{\pi/2}
\sin^{2}\theta\cos\phi d\theta d\phi},
\label{e7}
\end{equation}
\begin{equation}
\displaystyle
K(\varphi,\theta,\phi)=\Sigma_{\omega_{n}}
\frac{\bar{a}_{1}(\varphi,\theta,\phi)+\bar{a}_{2}(\varphi,\theta,\phi)
-\bar{a}_{1}(-\varphi,\theta,\phi)-\bar{a}_{2}(-\varphi,\theta,\phi)}
{2\Omega_{ns}},
\label{e8}
\end{equation}
\begin{equation}
\displaystyle
\bar{a}_{1}(\varphi,\theta,\phi)=\frac1i
\left\{\frac{A_{-}+\Gamma_{ns}B_{+}}{\Gamma_{ns}A_{-}-B_{+}}
\right\},
\hspace{10pt}
\displaystyle
\bar{a}_{2}(\varphi,\theta,\phi)=\frac1i
\left\{\frac{A_{+}+\Gamma_{ns}B_{-}}{\Gamma_{ns}A_{+}-B_{-}}
\right\},
\label{e3}
\end{equation}
\begin{equation}
A_{\pm}=e^{i\varphi/2}\gamma^{\ast}\Gamma_{np}(e^{-2XW}-1)\pm
e^{i\varphi}\Gamma_{ns}(e^{-2XW}+\Gamma_{np}^{2}),
\label{e4}
\end{equation}
\begin{equation}
B_{\pm}=e^{i\varphi/2}\gamma\Gamma_{ns}\Gamma_{np}(1-e^{-2XW})\pm
(1+\Gamma_{np}^{2}e^{-2XW}),
\label{e5}
\end{equation}
\begin{equation}
\displaystyle
\gamma=\frac{\Delta_{p}(T,\theta,\phi)}{\mid\Delta_{p}(T,\theta,\phi)\mid},
\displaystyle
\hspace{6pt}\Gamma_{ns(p)}=\sqrt{\frac{\Omega_{ns(p)}-\omega_{n}}
{\Omega_{ns(p)}+\omega_{n}}},
\end{equation}
\begin{equation}
\displaystyle
\Omega_{ns}=\sqrt{\omega_{n}^{2}+\mid\Delta_{s}(T)\mid^{2}},
\hspace{10pt}
\displaystyle
\Omega_{np}=
\sqrt{\omega_{n}^{2}+\mid\Delta_{p}(T,\theta,\phi)\mid^{2}}.
\label{e6n}
\end{equation}
Here $R_{N}$ denotes the normal resistance, with
$X=L/\xi_{p}$ and
$W=\Omega_{np}/(\mid\bar{\Delta}_{p}(0)\mid \sin\theta\cos\phi)$.
The coherence length in the triplet superconductor
$\xi_{p}$ is given as
$\xi_{p}=\hbar k_{F}/\mid m \bar{\Delta}_{p}(0)\mid$.
In the present paper, we select two kinds of unitary $p$-wave pair
potentials belonging to the two dimensional E$_{u}$ representation
which are expressed as
$\Delta_{\uparrow\downarrow}(T,\theta,\phi)=
\Delta_{p}(T,\theta,\phi)=
\bar{\Delta}_{p}(T)\sin\theta(\sin\phi+\cos\phi)$ and
$\bar{\Delta}_{p}(T)e^{i\phi}\sin\theta$.
Hereafter, we will call these as
E$_{u}$(U1) and E$_{u}$(U2) states, respectively.
The quantity $\gamma$ which is expressed by the internal phase of the
triplet superconductor
is given by $\gamma=1$ for E$_{u}$(U1),
and $\gamma=e^{i\phi}$ for E$_{u}$(U2).
By substituting these pair potentials in the above formulas,
we can obtain the Josephson current $J(\varphi)$ and $J_{C}(T)$.
\par
In the following calculation,
the transition temperatures $T_{p}$ and $T_{s}$
are set to be 1.0K and 8.1K, respectively, in order to
simulate Pb/Sr$_{2}$RuO$_{4}$/Pb junctions.
To express the two-dimensional features of the Fermi surface,
the quantity $\delta$ is chosen as $0.1\pi$.
The temperature dependence of $\Delta_{s}(T)$ and
$\bar{\Delta}_{p}(T)$ are assumed to obey the BCS relation.
Figures 2 and 3 show the calculated $J_{C}(T)$'s
for E$_{u}$(U1) and E$_{u}$(U2), respectively.
With the decrease of the $T$,
the characteristic temperature where
$J_{C}(T)$ begins to decrease is given by $T_{p}$.
With the further decrease of $T$,
$J_{C}(T)$ increases again.
Apparently these features are quite distinct from
those for conventional $s$-wave superconductors.
To understand this behavior intuitively, it is useful to present
analytical forms of $K(\varphi,\theta,\phi)$
in the two limiting cases.
For $T > T_{p}$, where
$\bar{\Delta}_{p}(T)=0$, is satisfied,
$K(\varphi,\theta,\phi)$ is reduced to be
\begin{equation}
\displaystyle
K(\varphi,\theta,\phi)=\Sigma_{\omega_{n}}\frac{1}{2\Omega_{ns}}
\left\{
\frac{4\Gamma_{ns}(1+\Gamma_{ns}^{2})e^{-2XW}\sin\varphi}
{\mid \Gamma_{ns}^{2}e^{-2XW} e^{i\varphi}+1\mid^{2}}
\right\},
\label{e9}
\end{equation}
which reproduces that  of in an ordinary $SNS$ junction \cite{ishii}.
While for the case $L\rightarrow\infty$,
$K(\varphi,\theta,\phi)$ is
calculated as
\begin{equation}
\displaystyle
K(\varphi,\theta,\phi)=\Sigma_{\omega_{n}}\frac{1}{2\Omega_{ns}}
\left\{\frac{-2\gamma^{\ast}\Gamma_{ns}(1+\Gamma_{ns}^{2})
\Gamma_{np}^{2}\sin\varphi}{(1-{\gamma^{\ast}}^{2}\Gamma_{ns}^{2}
\Gamma_{np}^{2})^{2}+4{\gamma^{\ast}}^{2}\Gamma_{ns}^{2}
\Gamma_{np}^{2}\sin^{2}(\varphi/2)}\right\}.
\label{e10}
\end{equation}
This result coincides with that of the
s-wave / $p$-wave junction \cite{yip}.
It should be emphasized
that a contribution to  $J(\varphi)$
from the former (referred to as $s/s$-coupling)
is positive and the latter is negative (referred to as $s/p$-coupling)
for $0<\varphi<\pi$.
This means that
the junction changes to so-called "$\pi$-junction"
when the $s/p$-coupling becomes dominant compared
to the $s/s$-coupling.
Since these two coupling terms have different $T$ and $L$ 
dependence,
the competition of these two determines the
behavior of the junction.
\par
Next, we will examine the thickness ($L$) dependence
of the Josephson current.
For sufficiently small $L$ ($L \ll \xi_{p}$),
$J(\varphi)$  remains positive independent of temperatures
because of the
strong $s/s$-coupling (see upper panel of Fig. 4).
When $L \approx \xi_{p}$,
the junction is also governed by the $s/s$-coupling
for $T>T_{p}$.
% or $L>\xi_{p}$,
However, with the decrease of temperatures, 
the contribution from $s/p$-coupling term [Eq. (\ref{e10})]
is rapidly enhanced. 
The sign changes in $J(\varphi)$ occurs when the $s/p$-coupling
becomes dominant at sufficiently low temperature.
At temperature just below $T_{p}$,
a crossover occurs between these two terms.
This is the origin of the non-monotonous
temperature dependence of $J_{C}(T)$.
While for $L \gg \xi_{p}$,
the $s/s$-coupling at $T>T_{p}$ is weakened because
it is governed by the factor $\exp[-LT/(\hbar v_{F})]$.
As the temperature is lowered below $T_{p}$,
the junction  easily transits to the $\pi$-junction \cite{Ges}.
Then the region, where $dJ_{C}(T)/dT >0$ is satisfied,
becomes narrower with the increase of $L$.
Thus the non-monotonous temperature
dependence is reduced and becomes undetectable.
\par
In this paper,
we present a theory of the Josephson current in a 
$s/p/s$ junction and obtain the non-monotonous temperature dependence of
$J_{C}(T)$.
This effect originates from the  competition between
the positive contribution of $K(\varphi,\theta,\phi)$
to $J(\varphi)$ due to the conventional $s/s$-coupling and
the negative one due to the  $s/p$-coupling.
The present effect is quite distinct from
the non-monotonous $J_{C}(T)$ predicted
in $d$-wave superconductor junction
\cite{tanaka,barash}
in the point that the present effect is insensitive
to the orientation of the junction.
Although we have assumed unitary state for Sr$_{2}$RuO$_{4}$,
it is also possible to perform the similar calculations for
several non-unitary pair potentials \cite{sigrist,machida,masashi}.
However, only the monotonous increase of $J_{C}(T)$ is expected
because of the absence of the Josephson coupling
between $s$-wave and these non-unitary pair potentials.
It is important to note that the present result
is consistent with the
recent measurements of Josephson current in
Pb/Sr$_{2}$RuO$_{4}$/Pb junction \cite{jin}.
We believe that the consistency strongly supports the unitary
triplet $p$-wave paring states in Sr$_{2}$RuO$_{4}$.
Since the  difference of the parity
plays an essential role in this effect,
the present trilayer junction configuration can be applied to
a novel phase-sensitive test for the unconventional pairing states.
For example, if E$_{u1}$ state \cite{sauls}
is also realized in UPt$_{3}$,
similar effect will be detected in Pb/UPt$_{3}$/Pb junctions.
%We hope our theory becomes useful to determine the symmetry
%of the pair potential in superconducting Sr$_{2}$RuO$_{4}$.
Throughout this paper,
we ignored the  barrier potential at the interface for the simplicity.
However, the essence of the physics will not be changed
even if we will take into account of this effect.
Details will be presented in the forthcoming paper.
\par
\vspace{0.5cm}
%======Acknowledgment=============================
We would like to thank to Y.~Maeno
for stimulating this topic
and for showing their results prior to
publications.
One of the author (S. K.) would like to thank to M. R. Beasley
for fruitful discussion.
This work is supported by a Grant-in-Aid for Scientific
Research in Priority Areas
''Anomalous metallic state near the Mott transition''
and ''Nissan Science Foundation''.
The computational aspect of this work has been done for the
facilities of the Supercomputer Center, Institute for Solid State Physics,
University of Tokyo and the Computer Center, Institute for Molecular
Science, Okazaki National Research Institute.
%================================================

\begin{figure}
\begin{center}
\epsfile{file=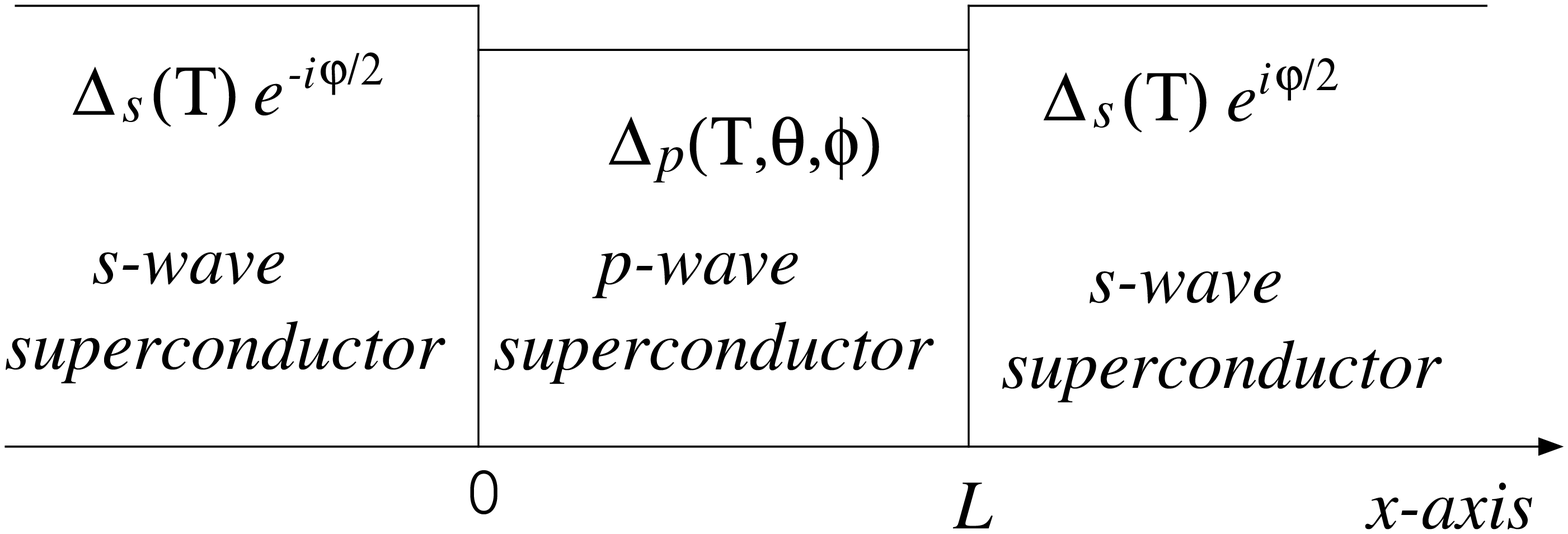,height=6cm,width=11cm,scale=1}
\end{center}
\caption{Schematic illustration  of $s$-wave superconductor /
$p$-wave superconductor / $s$-wave superconductor
($s/p/s$) junction.}
\label{f1}
\end{figure}

\begin{figure}
\begin{center}
\epsfile{file=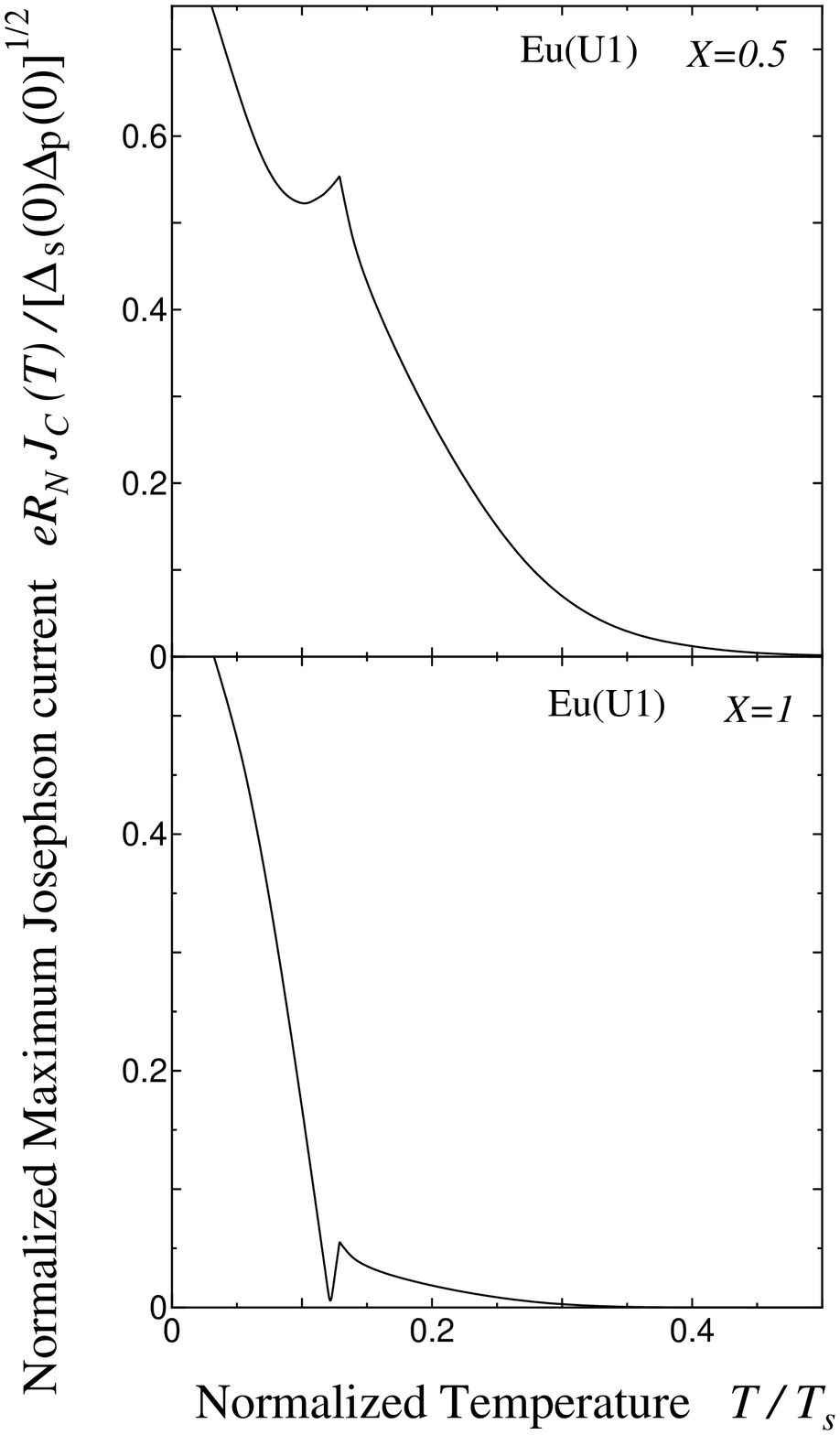,height=8cm,width=5cm,scale=1}
\end{center}
\caption{Normalized critical Josephson current as a function of
normalized temperature for $E_{u}$(U1)  state. $T_{p}/T_{s}=0.13$.
$X$ is 0.5 and 1 as indicated in the figure. }
\label{f2}
\end{figure}

\begin{figure}
\begin{center}
\epsfile{file=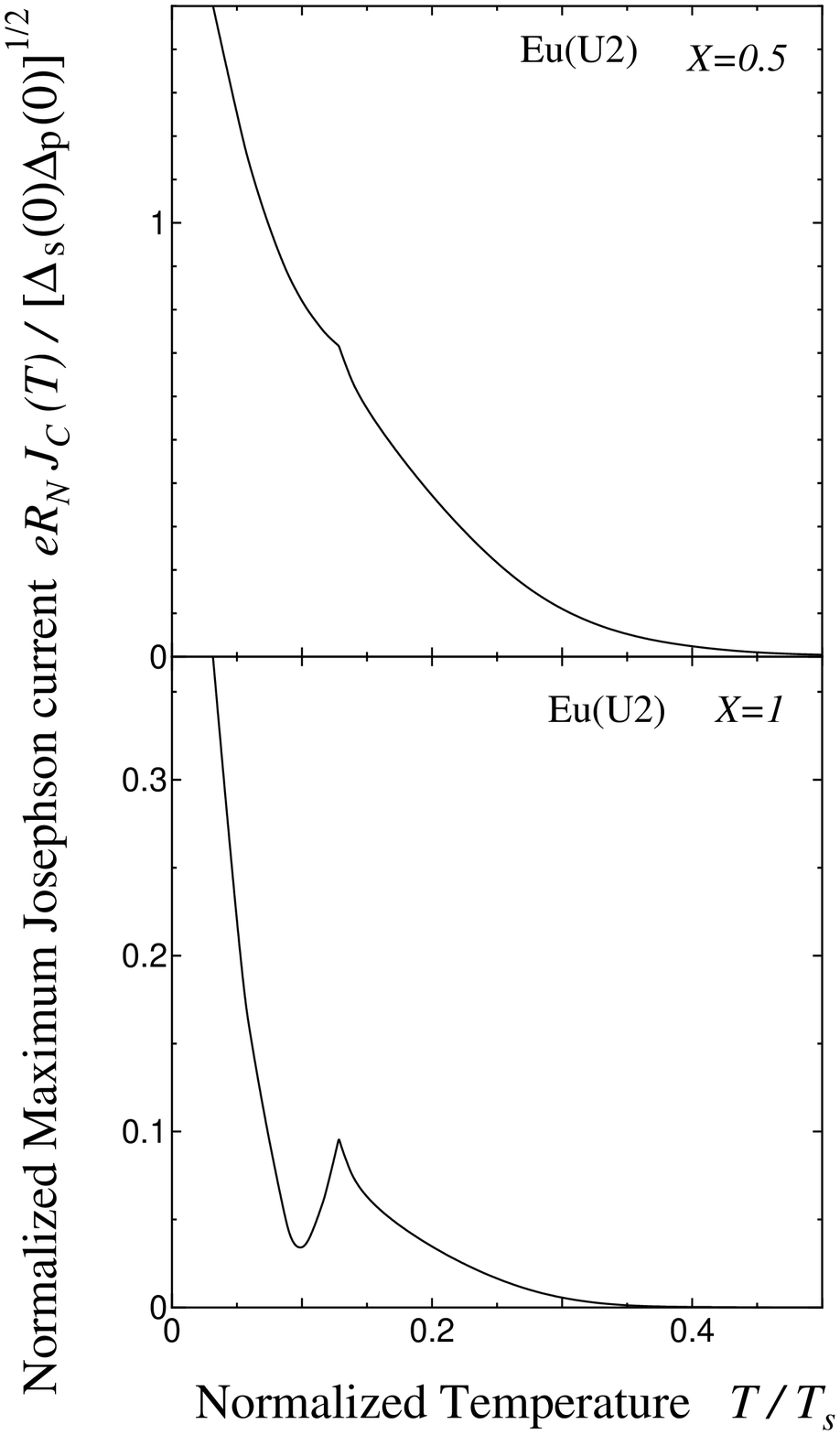,height=8cm,width=5cm,scale=1}
\end{center}
\caption{Normalized critical Josephson current as a function of
normalized temperature for $E_{u}$(U2)  state. $T_{p}/T_{s}=0.13$.
$X$ is 0.5 and 1 as indicated in the figure. }
\label{f3}
\end{figure}

\begin{figure}
\begin{center}
\epsfile{file=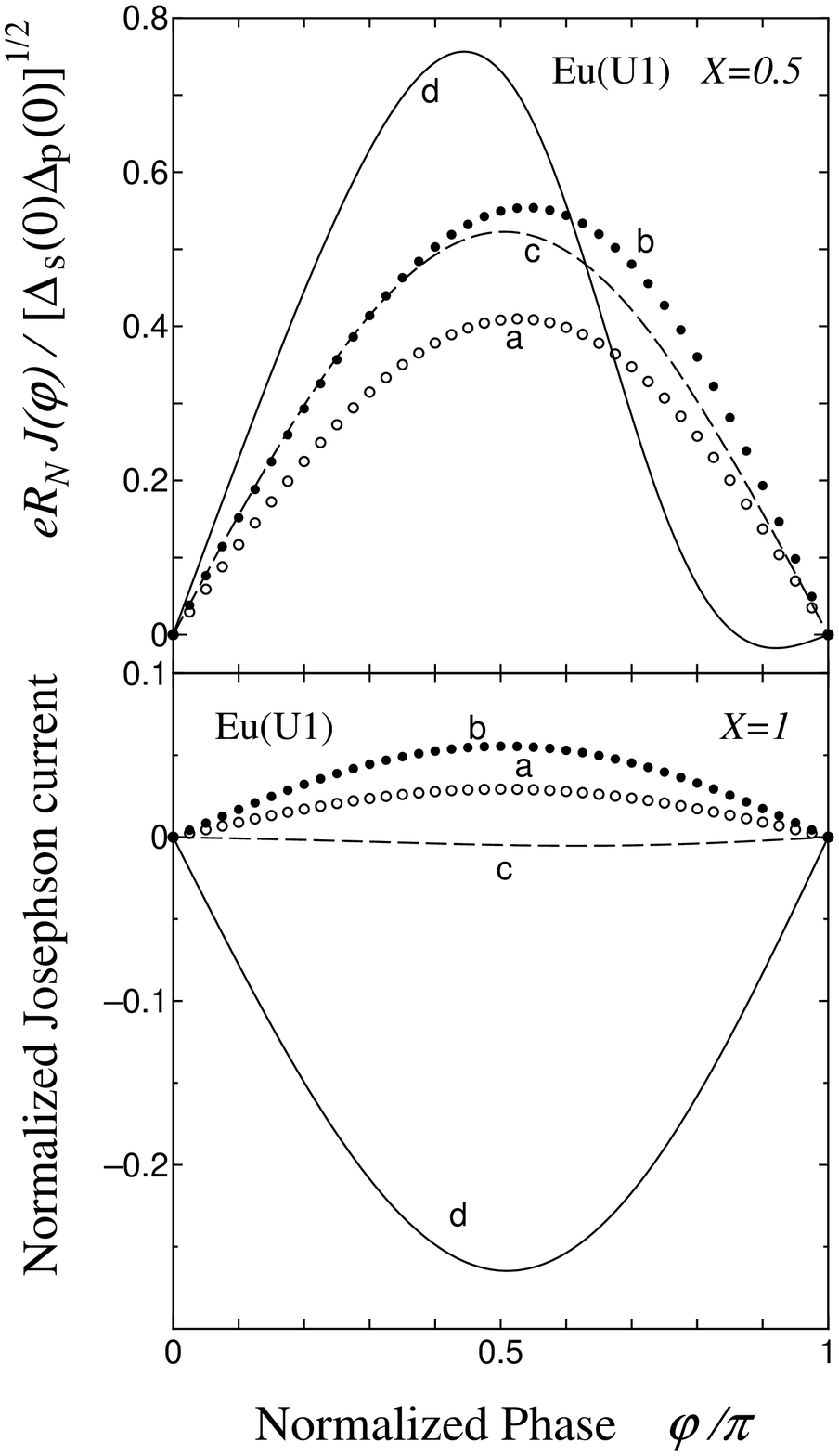,height=8cm,width=5cm,scale=1}
\end{center}
\caption{Normalized Josephson current as a function of
the phase difference $\varphi$ for $E_{u}$(U1)  state.
$T_{p}/T_{s}=0.13$.
$X$ is 0.5 and 1 as indicated in the figure.
For $X=0.5$, a:$T/T_{s}=0.15$, b:$T/T_{s}=0.13$, c:$T/T_{s}=0.11$,
and d:$T/T_{s}=0.03$.
For $X=1$, a:$T/T_{s}=0.15$, b:$T/T_{s}=0.13$, c:$T/T_{s}=0.12$,
and d:$T/T_{s}=0.09$.}
\label{f4}
\end{figure}

\end{document}